# Dirac-Surface-State Modulated Spin Dynamics in a Ferrimagnetic Insulator at Room Temperature

Chi Tang[1†], Qi Song[2, 8†], Cui-Zu Chang[3, 4], Yadong Xu[1], Yuichi Ohnuma[5], Mamoru Matsuo[5, 6], Yawen Liu[1], Wei Yuan[2, 8], Yunyan Yao[2, 8], Jagadeesh S. Moodera[3, 7], Sadamichi Maekawa[5], Wei Han[2, 8] and Jing Shi[1*]

[1]Department of Physics & Astronomy, University of California, Riverside, Riverside, CA 92521, USA
[2]International Center for Quantum Materials, School of Physics, Peking University, Beijing 100871, China
[3]Francis Bitter Magnet Laboratory, Massachusetts Institute of Technology, Cambridge, MA 02139, USA
[4]Department of Physics, The Pennsylvania State University, University Park, PA 16802, USA
[5]Advanced Science Research Center, Japan Atomic Energy Agency, Tokai 319-1195, Ibaraki, Japan
[6]Advanced Institute for Materials Research, Tohoku University, Sendai 980-8577, Miyagi, Japan
[7]Department of Physics, Massachusetts Institute of Technology, Cambridge, MA 02139, USA
[8]Collaborative Innovation Center of Quantum Matter, Beijing 100871, China

[†]These are co-first authors.

*Correspondence to: jing.shi@ucr.edu.

**Abstract:** This work demonstrates dramatically modified spin dynamics of magnetic insulator (MI) by the spin-momentum locked Dirac surface states of the adjacent topological insulator (TI) which can be harnessed for spintronic applications. As the Bi-concentration $x$ is systematically tuned in 5 nm thick $(Bi_xSb_{1-x})_2Te_3$ TI film, the weight of the surface relative to bulk states peaks at $x = 0.32$ when the chemical potential approaches the Dirac point. At this concentration, the Gilbert damping constant of the precessing magnetization in 10 nm thick $Y_3Fe_5O_{12}$ MI film in the MI/TI heterostructures is enhanced by an order of magnitude, the largest among all concentrations. In addition, the MI acquires additional strong magnetic anisotropy that favors the in-plane orientation with similar Bi-concentration dependence. These extraordinary effects of the Dirac surface states distinguish TI from other materials such as heavy metals in modulating spin dynamics of the neighboring magnetic layer.

# INTRODUCTION

Topological insulators (TI) are a new state of quantum matter with unique spin and charge properties owing to the non-trivial band topology and strong spin-orbit coupling (*1*). These properties can lead to a variety of exotic phenomena including topological magneto-electric effects (*2*), quantum anomalous Hall effect (*3*), image magnetic monopoles (*4*), etc. A remarkable feature that profoundly affects spin and charge transport in TI is that electrons in Dirac surface states have their spin locked orthogonally to their momenta in two dimensions (*5, 6*). Various spin and charge transport effects of such spin-momentum locking have been observed in spin valves with TI (*7*), spin Seebeck effect (*8*), inverse Edelstein effect (*9-13*) and spin-orbit torque switching (*14, 15*). Clearly, strong coupling between electron spin and translational degrees of freedom can be exploited as an efficient way of manipulating spins and vice versa, which are essential for spintronics.

Devices revealing the aforementioned effects are often constructed in heterostructures containing TI and a ferromagnetic layer which serves as either a pure spin current source or a spin detector. Any spin-dependent effect on transport properties is conveniently measured through the metallic surface states. In the same heterostructures, the reverse effect, i.e. the effect of the spin-momentum locked Dirac surface states on the spin dynamics of the magnetic layer, has not yet been systematically studied. It is known that a thin heavy metal such as Pt or W in contact with a magnetic material can cause broadening of the ferromagnetic resonance (FMR) linewidth, which is mainly attributed to the excess flow of spin current due to the spin pumping effect (*16, 17*); however, the linewidth broadening is generally insignificant (~10 Oe) (*18, 19*). In this regime, the effect is quantitatively described by the momentum sum of the imaginary part of the dynamical transverse spin susceptibility (*16*) or alternatively the spin-mixing conductance at the interface (*17*).

In this work, we investigate the room temperature spin dynamics of yttrium iron garnet ($Y_3Fe_5O_{12}$ or YIG), a ferrimagnetic insulator, in heterostructures containing $(Bi_xSb_{1-x})_2Te_3$, a TI material. By systematically tuning the chemical potential of the TI with a fixed thickness, i.e. five quintuple layers (QL) or ~ 5 nm, via varying Bi concentration $x$, we control the weight of the surface states relative to the bulk states (*20*). An alternative way of controlling the surface state weight is to use electrostatic gating which requires extensive materials work. In TI layers dominated by surface states, the direction of spins pumped into the Dirac surface states by FMR has to be locked in the plane. On the other hand, the spins in TI and those of YIG are exchange coupled to each other at the interface. As a result, the direction of the precessing spins in YIG is forced to be aligned in the plane (Fig. 1(a)). Consequently, spin pumping into Dirac surface states is expected to give rise to strong damping of the precessing magnetization. Indeed, we observe unprecedentedly large FMR linewidth broadening of YIG (up to 111 Oe at 9.6 GHz) when the chemical potential of the TI is tuned close to the Dirac point which corresponds to a 15 times larger Gilbert damping constant. This dramatic enhancement in damping is accompanied by an anomalously large increase (~ 100%) in the easy-plane anisotropy of the YIG layer.

## RESULTS

We choose YIG for the magnetic constituent in our heterostructures for several reasons to be described in the Materials and Methods section. In this study, we have fabricated seven heterostructure samples with various Bi concentrations, $x$, ranging from 0 to 1. $x = 0.32$ sample is most insulating as indicated by the largest resistance at 300 K (Fig. 1(b)) and the largest negative slope in its temperature dependence (Fig. 1(c)), suggesting that the chemical potential is located close to the Dirac point. With $x$ deviating from 0.32, the chemical potential is tuned away from the Dirac point (20). In $Bi_2Te_3$ ($x = 1$) and $Sb_2Te_3$ ($x = 0$), the chemical potential is located in the bulk conduction and valence bands, respectively.

We measure FMR spectra with both cavity and broad-brand co-planar waveguide FMR setups. Prior to TI layer deposition, cavity FMR measurements are performed on all 10 nm thick YIG films at room temperature using a Bruker 9.6 GHz X-band EMX EPR spectrometer. Fig. 2(b) (red curve) shows a FMR derivative absorption spectrum of a representative YIG film with an in-plane magnetic field, which can be well fitted with a single Lorentzian derivative. The peak-to-peak linewidth $\Delta H_{pp}$ of 7.0 Oe and resonance field $H_{res}$ of 2434 Oe are obtained from the fitting. Such a narrow linewidth indicates high YIG film quality. The mean values of both $\Delta H_{pp}$ and $H_{res}$ for all seven samples are 10.2 ± 3.1 Oe and 2394.5 ± 40.2 Oe, respectively. When FMR is performed as a function of the polar angles $\theta_H$ (defined in Fig. 2(a)), the $\theta_H$-dependence of both quantities shows very small variations for all seven YIG samples (fig. S2), indicating a tight control over the YIG film quality.

To study the effect of the TI Dirac surface states on YIG spin dynamics, we compare the YIG FMR spectra taken before and after the TI growth. Fig. 2(b) shows a direct comparison between the two representative FMR spectra taken with in-plane fields before and after the growth of a 5 QL $Sb_2Te_3$ ($x = 0$) film. Two distinct differences stand out. First, $H_{res}$ is shifted to a lower field by 207 Oe, i.e. from 2434 Oe to 2227 Oe, indicating a large effect on magnetic anisotropy upon adding the 5 QL $Sb_2Te_3$. Second, the FMR linewidth $\Delta H_{pp}$ is broadened by seven times. To more accurately determine the effective anisotropy field change, we measure FMR of both YIG/$Sb_2Te_3$ and a YIG reference sample as a function of polar angle $\theta_H$ (21, 22). Fig. 2(c) shows the spectra at selected polar angles between the in-plane ($\theta_H = 90°$) and out-of-plane ($\theta_H = 0°$) magnetic field orientations. The YIG reference sample has relatively narrow FMR linewidth for all polar angles. In the meantime, $H_{res}$ decreases monotonically from ~5560 Oe for $H$ out-of-plane to ~ 2434 Oe for $H$ in-plane, consistent with the behavior of nanometer thick YIG films with easy-plane magnetic anisotropy (23). With 5 QL $Sb_2Te_3$ on top, however, dramatic differences can be readily identified as shown in the top panel of Fig. 2(c): significant broadening of $\Delta H_{pp}$ and large shift in $H_{res}$. $H_{res}$ shift occurs at all angles; hence, the overall $H_{res}$ range is greatly expanded, i.e. between 2227 Oe and 6000 Oe. Such marked effects are not seen in heterostructures containing thin heavy metal layers such as Pt (18).

The differences in both $\Delta H_{pp}$ and $H_{res}$ caused by the 5 QL $Sb_2Te_3$ suggest their origin in TI's band structure. In order to study the effects of the TI Dirac surface states, we compare the FMR results in all seven samples in which the surface-to-bulk ratio systematically varies. We first study the effect of varying $x$ on the easy-plane magnetic anisotropy. We plot $H_{res}$ as a function of $\theta_H$ in Fig. 3(a) for all seven YIG/$(Bi_xSb_{1-x})_2Te_3$ samples plus a YIG reference film. The angular dependence data can be fitted reasonably well with the shape anisotropy plus a uniaxial anisotropy term, both in the form of $\cos^2\theta$. As is routinely done in literature (21), we solve three transcendental equations numerically and seek the least-square fitting results. Note that if the uniaxial anisotropy term is negative, it simply represents easy-plane anisotropy. We find that it is indeed negative, i.e. there is additional easy-plane anisotropy in all seven samples. By fitting the polar angle dependence, we obtain the two best-fit parameters, $4\pi M_{eff}$ and the $g$-factor for each sample. The effective anisotropy field is defined as $4\pi M_{eff} = 4\pi M_s + H_{an}$, where $4\pi M_s$ and $H_{an}$ denote the demagnetizing field and an effective easy-plane (positive) magnetic anisotropy field, respectively (24). The best-fit parameters are summarized in table. S1. The extracted $4\pi M_{eff}$ is plotted in Fig. 3(b). Since the demagnetizing field $4\pi M_s$ is about 1780 Oe, $4\pi M_{eff}$ is clearly larger than $4\pi M_s$ in all samples. Furthermore, $4\pi M_{eff}$ depends on $x$. For the two most metallic TI films, i.e. $x = 0$ and $x = 1$, $4\pi M_{eff}$ is increased by 420 and 460 Oe, which accounts for 23% and 26% of its demagnetizing field $4\pi M_s$, respectively. In comparison, the corresponding increase is only 5% in YIG/Pt (fig. S3(a)). More interestingly, as the chemical potential is tuned into the band gap, i.e. surface states becoming dominant, $4\pi M_{eff}$ increases further and peaks in the most insulating sample ($x = 0.32$), reaching 3800 Oe. This increase represents nearly a 100% enhancement over the YIG demagnetizing field $4\pi M_s$.

A common origin of enhanced magnetic anisotropy in thin films is related to the interface strain. In $(Bi_xSb_{1-x})_2Te_3$, the interaction between the neighboring Te-Bi/Sb-Te-Bi/Sb-Te quintuple layers is of the van der Waals type. Between TIG and TI, there is no epitaxial relation due to widely different lattice structures; therefore, the strain and strain-induced anisotropy are expected to be small at the YIG-TI interface for all samples. Another possibility of the enhanced $4\pi M_{eff}$ is an increased demagnetizing field. If part of TI becomes ferromagnetic, it can in principle cause an increase in $4\pi M_s$. We exclude this possibility with the following arguments. First, such a proximity induced moment, if exists, can only come from a few atomic layers at the interface and is clearly too small to account for the observed 100% increase. Our magnetization measurements do not support this possibility either (fig. S5). The same magnetometry results do not show any clear Bi-concentration dependence within experimental uncertainty. More importantly, the proximity effect in YIG/TI heterostructures occurs at much lower temperatures (< 150 K) as reported in previous studies (8, 25). Additionally, the $T_c$ of the induced ferromagnetism in TI was found to be uncorrected with TI's chemical potential position (25). In our data, the $4\pi M_{eff}$ enhancement follows the same trend as the resistivity as shown in Fig. 1(c). From these analyses, we conclude that the enhanced effective anisotropy originates from the Dirac surface states.

The cavity-FMR measurements have already indicated anomalously broadened FMR linewidth at a particular microwave frequency. In order to extract the Gilbert damping constant $\alpha$, we perform broadband FMR measurements using a coplanar waveguide setup for all YIG/TI samples up to 12 GHz. Representative transmission data $S_{21}$ in Fig. 4(a) show the FMR absorption of YIG/$Sb_2Te_3$ with several frequencies. Both FMR resonance field shift and linewidth broadening display the same trend in Fig. 4(b). The half width at half maximum, $\Delta H = \sqrt{3}\Delta H_{pp}/2$, is extracted by fitting a Lorentzian function to each $S_{21}$ spectrum up to 12 GHz, and then plotted in Fig. 4(c) for all samples. A linear relation between $\Delta H$ and frequency is observed, and $\alpha$ can be calculated by(24)

$$\Delta H = \frac{2\pi}{\gamma}\alpha f + \Delta H_0, \qquad (1)$$

where $\gamma$ and $\Delta H_0$ are the gyromagnetic ratio and the inhomogeneity linewidth broadening, respectively. Fig. 4(d) shows $\alpha$ vs. $x$ for all samples. Interestingly, $\alpha$ peaks at $x = 0.32$ as well, i.e. in the most insulating sample, similar to the resistivity and the effective anisotropy field. Compared to the two most metallic samples with $\alpha = 8.0 \times 10^{-3}$ for $x = 0$ (or $Sb_2Te_3$) and $\alpha = 4.5 \times 10^{-3}$ for $x = 1$ (or $Bi_2Te_3$), $\alpha$ reaches the maximum value of $2.2 \times 10^{-2}$ for $x = 0.32$, which is an order of magnitude larger than that of the bare YIG films (average value of $1.5 \times 10^{-3}$). In comparison, $\alpha$ in YIG/Pt is only twice as large as that in the YIG reference, as shown in fig. S4(b). Additionally, $\alpha$ shows the same trend as that of the resistivity and effective anisotropy. These facts suggest a common origin, i.e. the special band structures of the TI surface states rather than the spin-orbit coupling of the constituent elements. The latter effect would imply a monotonically increasing trend as more Bi atoms are incorporated.

## DISCUSSION

We now show that these three effects are actually connected and given by the spin-momentum locking properties of the TI Dirac surface states. The spins ($\vec{\sigma}$) of TI and the spins ($\vec{S}$) of YIG are coupled by the exchange interaction expressed as (26, 27), $H = J_{sd}\vec{\sigma} \cdot \vec{S}$, with $J_{sd}$ being the exchange constant at the interface. We note that the spins in TI lie in the plane due to the spin-momentum locking. Therefore, the spins in YIG are pulled towards the plane as well. This pulling effect induces the easy-plane anisotropy given by

$$E_{an} = -\frac{1}{2}(\chi_{//} - \chi_{\perp})M^2 \qquad (2)$$

where $\chi_{//}$ and $\chi_{\perp}$ are the in-plane and out-of-plane components of the spin susceptibility in TI, respectively, and $M$ is the magnetization of YIG. In particular, when the chemical potential is close to the

Dirac point, $\chi_\perp$ is strongly suppressed due to the gap(28) and, thus, $\chi_{//} \gg \chi_\perp$. TI's spin susceptibility gives rise to the easy-plane magnetic anisotropy field by $H_{an} = -\frac{\partial E_{an}}{\partial M}$.

The same interfacial exchange interaction also affects spin pumping (16); the spin precession of YIG in FMR results in the motion of spins in TI. The induced spin current ($I_s$) in TI is expressed as

$$I_s \propto J_{sd}^2 I_m \chi^{+-}(\omega) \qquad (3)$$

where $\chi^{+-}(\omega)$ is the transverse component in TI to the magnetization direction of YIG, and $\omega$ is the FMR frequency. We note that since the spin Seebeck effect is due to the spin pumping by heat (26, 27), it is also given by $\chi^{+-}(\omega)$ but integrated over $\omega$ in the range of the thermal distribution of spin fluctuations. The susceptibility is calculated by taking into account the direct transition near the Fermi level because of the spin-momentum locking in TI (see SI for details). Since the resonance frequency (a few GHz) is much smaller than the energy gap of TI (~ 0.3 eV) (29), the direct transition is more effective when the Fermi energy is near the Dirac point. As a result, the spin pumping (the Gilbert damping) is significantly enhanced near the Dirac point. This model also explains enhanced spin Seebeck effect reported previously (8).

In summary, we have observed dramatic modifications of YIG spin dynamics by spin-momentum locked surfaces of a thin TI layer in high-quality YIG/TI heterostructures with different Bi/Sb ratios. The spin-momentum locking in TI provides not only a sensitive detection of the magnetic state in magnetic materials serving as a spin current source, but also an active way of manipulating ultrafast magnetization dynamics and magnetic anisotropy with the unique properties of the topological Dirac surface states, which offers exciting opportunities for potential spintronic applications.

## MATERIALS AND METHODS

**Choice of YIG**: We choose 10 nm thick YIG films as the MI layers in all MI/TI heterostructures. First, YIG in general has a very small Gilbert damping constant $\alpha$ ( ~ $3 \times 10^{-5}$ in crystals and ~ $10^{-3}$ in 10 nm thick YIG films); and the FMR linewidth is relatively narrow (~ 10 Oe at ~ 10 GHz for thin films). Therefore, small linewidth changes can be easily detected. Second, YIG films are prepared first with high temperatures (~ 800 °C) with pulsed laser deposition and rapid thermal annealing and the TI layers are grown at much lower temperatures (~230 °C) after with molecular beam epitaxy. This growth sequence and the large temperature difference prevent serious intermixing across the interface. In our heterostructures, YIG is atomically flat, which ensures the flat YIG-TI interface. Third, similar to our previous spin Seebeck effect study (8), here we conduct FMR measurements at room temperature which is well above that of the induced ferromagnetism in the TI surface layer. Consequently, the dynamic behavior of YIG is not affected by the induced ferromagnetism in TI (18).

**Heterostructure growth:** Thin YIG films are grown on epi-ready lattice-matched single crystal GGG (111) substrates via pulsed laser deposition. The detailed growth recipe of YIG films is described in a previous paper (*30*). To fabricate high-quality YIG/(Bi$_x$Sb$_{1-x}$)$_2$Te$_3$ heterostructures with various Bi concentrations (x = 0, 0.22, 0.27, 0.32, 0.46, 0.67 and 1 in this study), YIG (111) films are then transferred to an ultra-high vacuum molecular beam epitaxy (MBE) system with the base pressure better than 5×10$^{-10}$ Torr for TI growth. High-purity Bi (99.999%), Sb (99.9999%) and Te (99.9999%) are evaporated from Knudsen effusion cells. During the growth, the YIG substrate is kept at 230 °C and the growth rate of TI is ~0.2 QL/min. The heterostructure film is covered with a 5 nm Te protection layer before taken out of the MBE chamber for the FMR measurements.

**FMR measurements:** The polar angle dependent FMR measurements for all samples are performed using a Bruker 9.6 GHz X-band EMX EPR spectrometer. Samples can be rotated with respect to the static field direction from the in-plane to out-of-plane geometry with a protractor reading the angle precisely.

The Gilbert damping constant measurements are conducted by a broad-band FMR using coplanar waveguide setup. The forward amplitude of complex transmission coefficients ($S_{21}$) is recorded by the vector network analyzer (VNA, Agilent E5071C) connected to a straight-line coplanar waveguide (*31*). Sample is attached to the waveguide and the measurement is performed with the frequency sweeping from 1 GHz to 12 GHz at a fixed magnetic field which can be varied up to 4000 Oe.

## SUPPLEMENTARY MATERIALS

fig. S1. Crystal structure and surface morphology.
fig. S2. FMR resonance field and linewidth for all YIG films.
fig. S3. Effect of Pt on YIG resonance characteristics.
fig. S4. Comparison of FMR between YIG/TI and YIG/Pt.
fig. S5. Comparison of total effective anisotropy field and demagnetizing field.
fig. S6. High-resolution transmission electron microscope image of a representative YIG/TI sample.
fig. S7. TI surface state dispersion.
fig. S8. Direct transition of TI conduction electrons driven by spin pumping.
table S1. Two parameters (i.e. $4\pi M_{eff}$ and $\gamma/2\pi$) obtained from fitting.

**Acknowledgments**

We acknowledge the assistance from N. Samarth and J. Kally for the sample preparation and useful discussions with Z. Shi, J. Li, V. Ortiz and M. Aldosary.

**Funding:** This work was supported as part of the SHINES, an Energy Frontier Research Center funded by the U.S. Department of Energy, Office of Science, Basic Energy Sciences under Award No. SC0012670 (CT, YDX, YWL and JS). QS, WY, YYY and WH acknowledge the support of the National Basic Research Programs of China (973 Grants 2014CB920902 and 2015CB921104) and the National Natural Science Foundation of China (NSFC Grant 11574006). CZC and JSM acknowledge the support from NSF Grants No. DMR-1207469, ONR Grant No. N00014-16-1-2657, and the STC Center for Integrated Quantum Materials under NSF Grant No. DMR-1231319.

**Author contributions:** JS conceived and supervised the experiments. CT grew the YIG magnetic insulator thin films and performed cavity FMR measurements and data analysis with the help of YWL. QS performed the broadband FMR measurements and data analysis with the help of WY and YYY and the supervision of WH. YDX performed TEM sample preparation. CZC grew the topological insulator thin films on YIG to form heterostructures with JSM's supervision. YO, MM and SM did theoretical calculations. All authors participated in the preparation of the final manuscript.

**Competing interests:** The authors declare that they have no competing interests. Data and materials availability: All data needed to evaluate the conclusions in the paper are present in the paper and/or the Supplementary Materials. Additional data related to this paper may be requested from the authors.

**FIGURES AND TABLES**

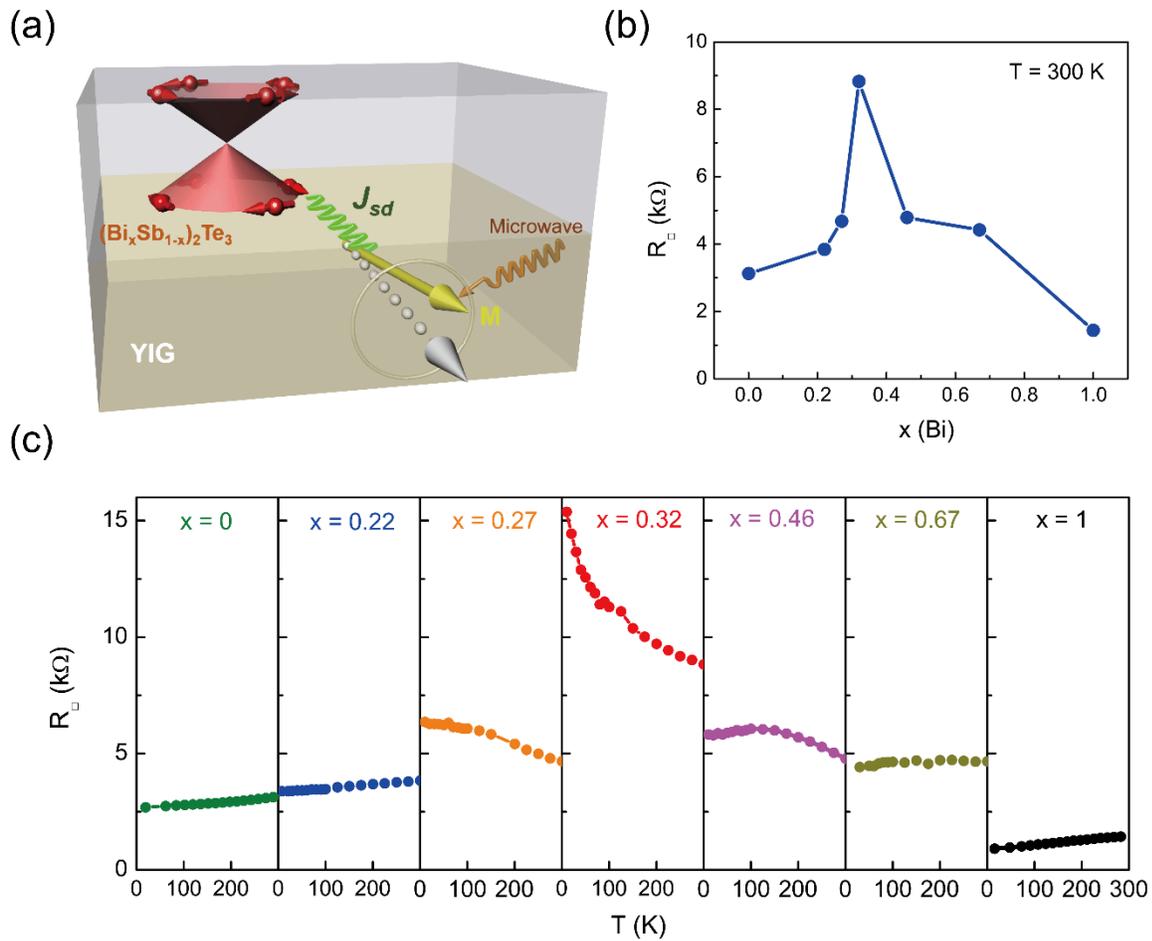

**Fig. 1. FMR measurement principle and TI properties.** (a) Schematic drawing of magnetization dynamics in YIG interfaced with TI in which the spin of the surface state electron is locked to momentum. (b) Room temperature sheet resistance of $(Bi_xSb_{1-x})_2Te_3$ with different Bi concentrations. (c) Temperature dependence of the sheet resistance of seven YIG (10 nm)/$(Bi_xSb_{1-x})_2Te_3$ (5 QL) heterostructures.

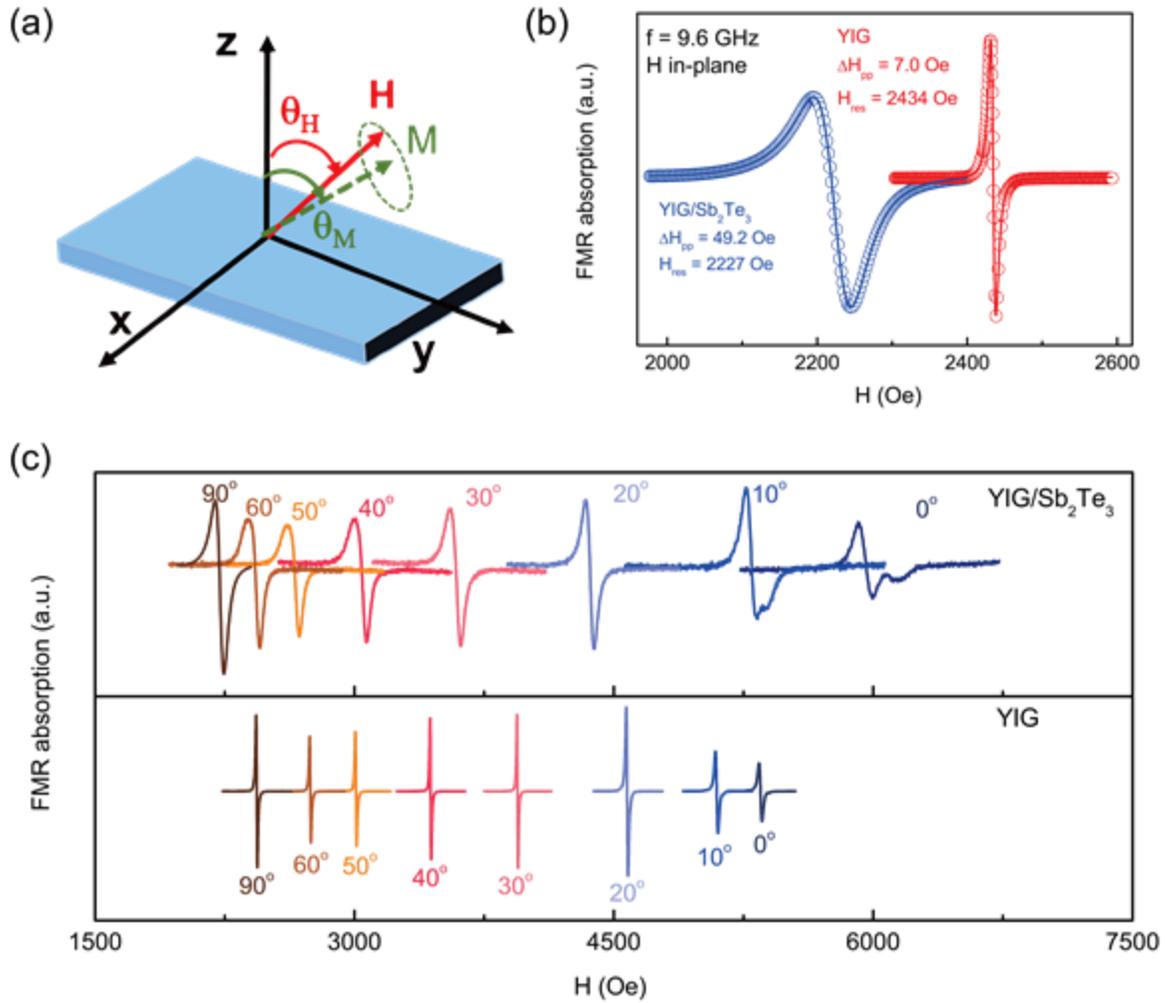

**Fig. 2. YIG FMR spectra with and without TI.** (a) Definition of polar angles, $\theta_H$ and $\theta_M$ in FMR measurements. (b) FMR derivative absorption spectra of YIG/Sb$_2$Te$_3$ and YIG reference sample at a frequency of 9.6 GHz with magnetic field applied in-plane ($\theta_H = 90°$). The solid lines are the best fits to extract the resonance field $H_{res}$ and peak-to-peak linewidth $\Delta H_{pp}$. (c) FMR derivative absorption spectra of YIG/Sb$_2$Te$_3$ and YIG reference sample with the polar angle $\theta_H$ ranging from 0°(out-of-plane) to 90° (in-plane) at 300 K. The extra peak-like feature on the high field side of the resonance at 0° and 10° is also observed some other samples, which could be caused by minor inhomogeneity change in YIG due to the presence of the TI layer.

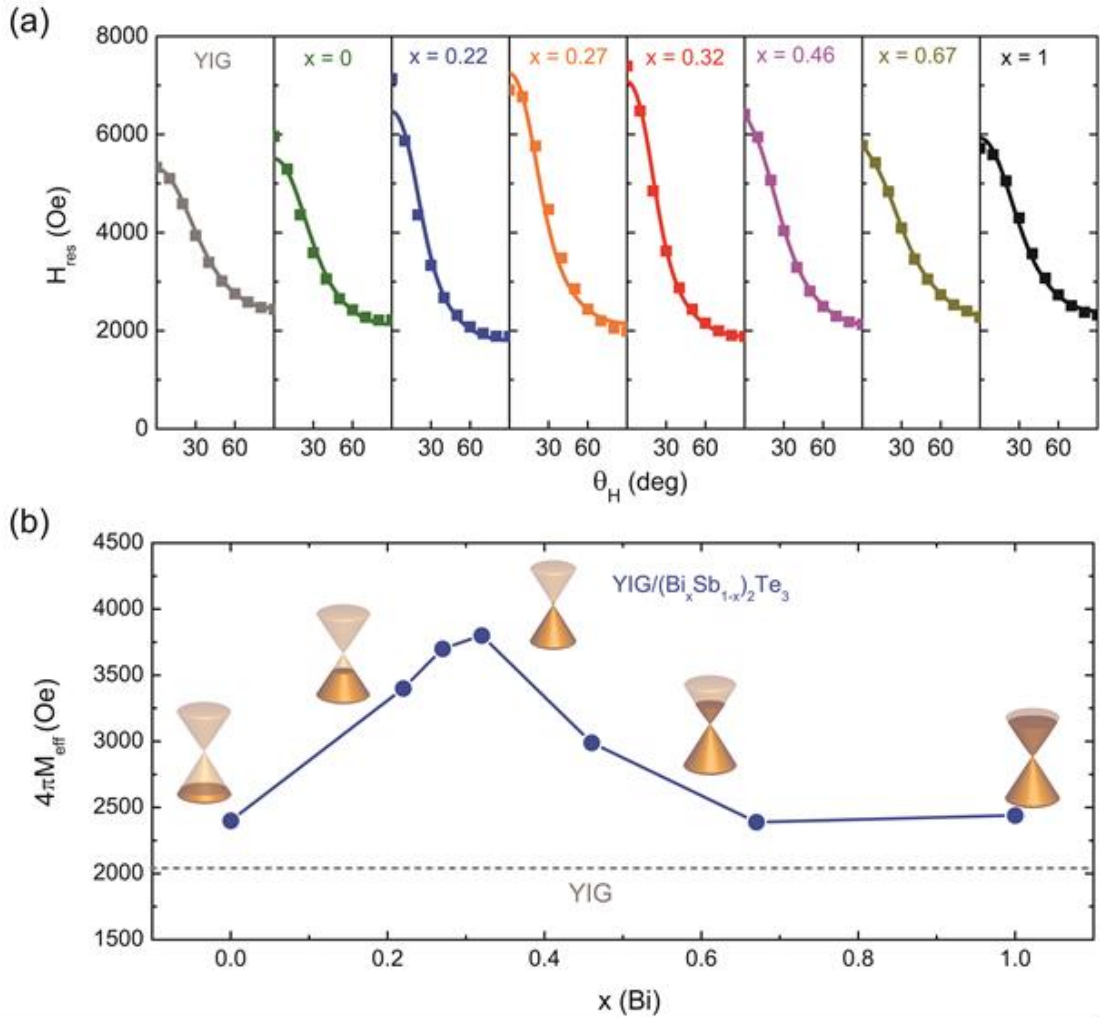

**Fig. 3. Extracted $4\pi M_{eff}$ from FMR polar angle dependence fitting.** (a) Polar angle $\theta_H$ dependence of FMR resonance field $H_{res}$ for all seven YIG (10 nm)/(Bi$_x$Sb$_{1-x}$)$_2$Te$_3$ (5 QL) samples and YIG reference sample. Solid curves are the best fits. (b) Bi-concentration dependence of extracted effective anisotropy field $4\pi M_{eff}$ obtained from fitting in (a) for all seven YIG (10 nm)/(Bi$_x$Sb$_{1-x}$)$_2$Te$_3$ (5 QL) samples. The black dashed line is the $4\pi M_{eff}$ value for the YIG reference sample.

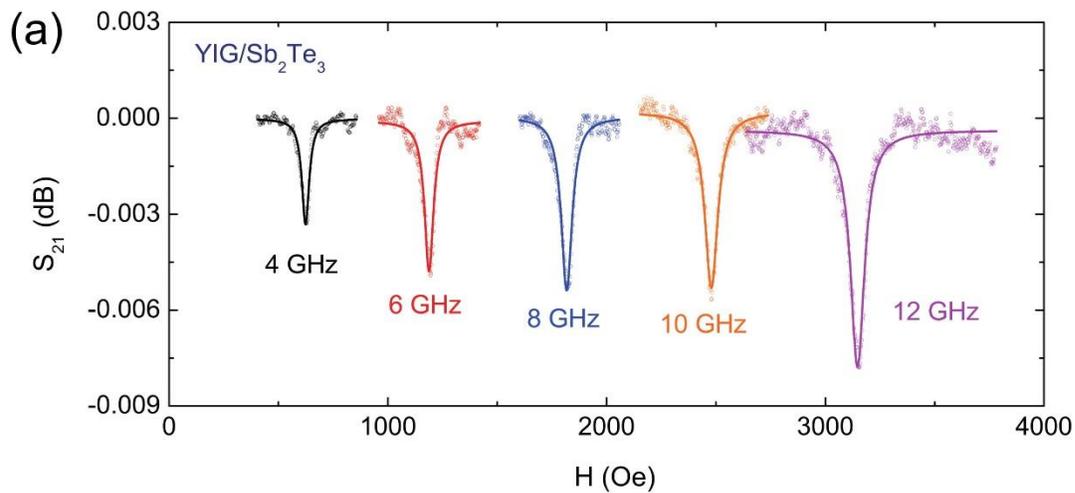
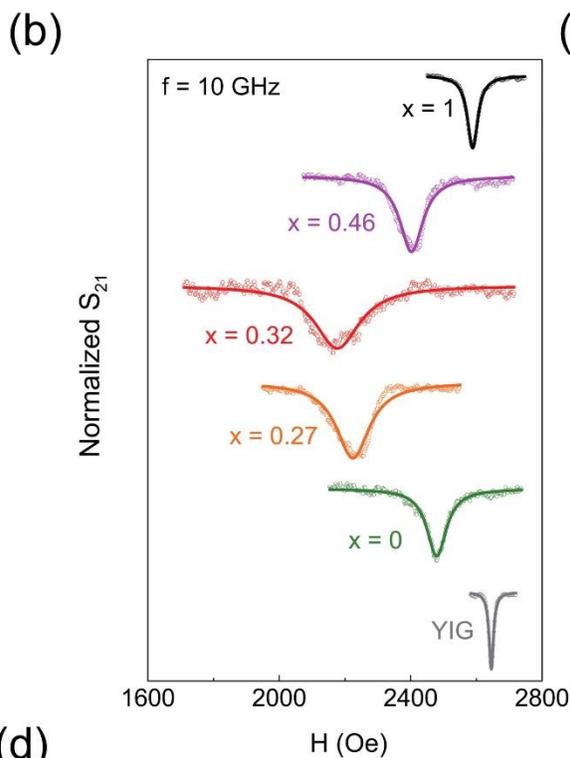
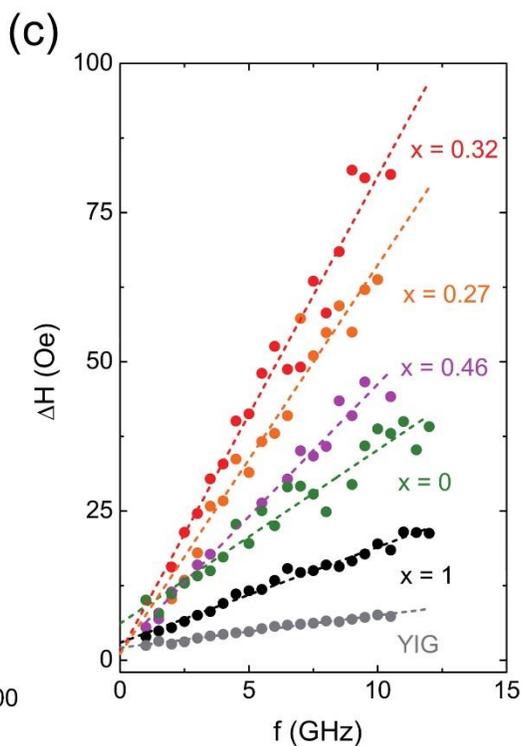
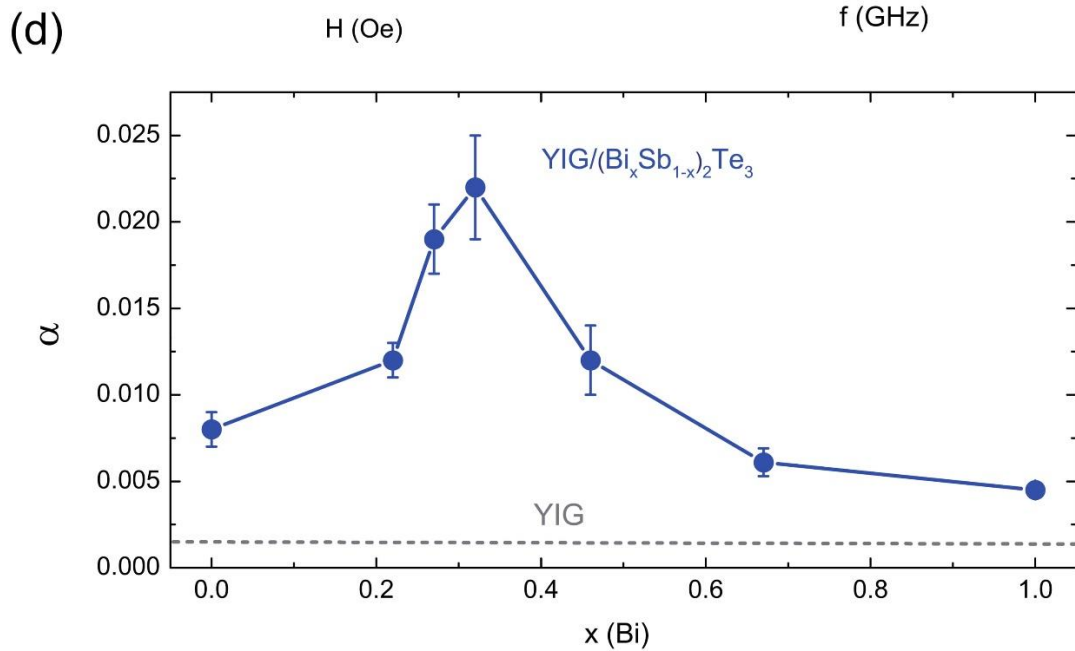

**Fig. 4. Extracted Gilbert damping from FMR linewidth fitting.** (a) FMR transmission spectra $S_{21}$ for YIG/Sb$_2$Te$_3$ for different chosen frequencies: 4, 6, 8, 10 and 12 GHz at 300 K after background subtraction. (b) Normalized FMR spectra $S_{21}$ at a fixed frequency of 10 GHz with an applied in-plane static field for YIG/(Bi$_x$Sb$_{1-x}$)$_2$Te$_3$ samples with different Bi concentrations and YIG reference sample. (c) Frequency dependence of FMR linewidth for all seven YIG/TI samples and YIG reference sample. The resonance peak height is reduced in samples with increased damping constant, which causes poor Lorentzian fitting and consequently large apparent noise in extracted linewidth. (d) Bi concentration ($x$)-dependence of the Gilbert damping constant $α$ extracted from the slope of the straight lines in (c).